\documentclass[12pt,a4paper,titlepage]{article}
\pdfoutput=1                          
\usepackage[utf8x]{inputenc}
\usepackage[english]{babel}

\usepackage{ucs}
\usepackage{amsmath}
\usepackage{amsfonts}
\usepackage{amssymb}
\usepackage{eucal}
\usepackage{mathrsfs}          
\usepackage[sans]{dsfont}     
\usepackage{slashed}
\usepackage[Symbol]{upgreek}  
\usepackage{array}            
\usepackage{cite}               
\usepackage[normalem]{ulem}             
\usepackage{booktabs}      
\usepackage{colortbl}         
\usepackage{fancyhdr}
\usepackage{multirow}
\usepackage{feynmf}           
\usepackage
{graphicx}

\usepackage{color}
\definecolor{grigio}{cmyk}{0,0,0,0.1}
\definecolor{rosa}{cmyk}{0,0.1,0.1,0.02}
\definecolor{rosino}{cmyk}{0,0.05,0.05,0.02}
\definecolor{rosas}{cmyk}{0,0.3,0.25,0.05}
\definecolor{celeste}{cmyk}{0.1,0,0,0.02}
\definecolor{giallino}{cmyk}{0,0,0.1,0.02}
\definecolor{rosso}{cmyk}{0,1,1,0.4}
\definecolor{rossos}{cmyk}{0,1,1,0.55}
\definecolor{rossoc}{cmyk}{0,1,1,0.2}
\definecolor{blu}{cmyk}{1,1,0,0.3}
\definecolor{blus}{cmyk}{1,1,0,0.5}
\definecolor{bluc}{cmyk}{1,1,0,0.1}
\definecolor{blucc}{cmyk}{0.7,0.5,0,0}
\definecolor{viola}{cmyk}{0,1,0,0.6}
\definecolor{viola2}{cmyk}{0,1,0.2,0.6}
\definecolor{verde}{cmyk}{0.92,0,0.59,0.25}
\definecolor{verdec}{cmyk}{0.92,0,0.59,0.15}
\definecolor{verdes}{cmyk}{0.92,0,0.59,0.4}
\definecolor{verdino}{cmyk}{0.12,0,0.09,0.02}
\definecolor{giallo}{cmyk}{0,0,1,0}
\definecolor{gialloverde}{cmyk}{0.44,0,0.74,0}
\definecolor{Titolo}{rgb}{0.752941176,0.576470588,0.992156863}
\definecolor{altro}{rgb}{0.094117647,0.650980392,0.643137255}
\definecolor{Peanuts}{rgb}{0.2, 0.4, 0.6}
\definecolor{Pean1}{rgb}{0.6, 0.8, 0.4}
\definecolor{BHO}{rgb}{0.2, 0.8, 1}
\definecolor{Daria}{rgb}{0, 0.9412, 0}
\definecolor{UniPi}{rgb}{0.2549, 0.4627, 0.6275}
\definecolor{UniPidue}{rgb}{0.3216, 0.5804, 0.7882}
\definecolor{rossoCP3}{cmyk}{0,.88,.77,.40}
\definecolor{verdeCP3}{rgb}{0.09765625, 0.57421875, 0.1015625}
\definecolor{bluCP3}{rgb}{0, 0.23, 0.67}
\definecolor{bluSaclay}{rgb}{0, 0.22, 0.70}

\usepackage{cancel}
\usepackage[normalem]{ulem}

\newcommand{\beq}{\begin{equation}}
\newcommand{\eeq}{\end{equation}}

\long\def\symbolfootnote[#1]#2{\begingroup\def\thefootnote{\fnsymbol{footnote}}\footnote[#1]{#2}\endgroup}

\ifx\pdfoutput\undefined
\usepackage[dvips,bookmarks]{hyperref}    
\else
\usepackage{hyperref}    
\fi
\hypersetup{colorlinks, bookmarksopen, bookmarksnumbered,
citecolor=verdeCP3, linkcolor=bluCP3, pdfstartview=FitH, urlcolor=rossoCP3}

\def\mhref#1{\href{mailto:#1}{#1}}        

\def\EDGES{{\sc EDGES}}

\begin{document}

\begin{titlepage}

\begin{flushright}
\scriptsize
\end{flushright}
\color{black}
\vspace{0.3cm}

\begin{center}
{\Huge{\color{blus}\bf 21-cm line Anomaly: A brief Status}\\[5mm]
\rule{0pt}{25pt}}
\end{center}
\begin{center}
{\sc Paolo Panci$\, {\color{bluCP3}^{1,2,3}}$\,\symbolfootnote[1]{\mhref{paolo.panci@unipi.it}}}
\end{center}

\begin{center}
\par \vskip .1in \noindent
{\it ${\color{bluCP3}^1} \,$\href{https://www.df.unipi.it}{Dipartimento di Fisica, Universit\`a di Pisa} and INFN,  Sezione di Pisa \\ Largo Pontecorvo 3, 56127 Pisa, Italy }
\\
  {\it ${\color{bluCP3}^3} \, $\href{https://www.lngs.infn.it}{Laboratori Nazionali del Gran Sasso}, INFN-LNGS \\ Via G.~Acitelli, 22, I-67100 Assergi (AQ), Italy}
  \\
{\it ${\color{bluCP3}^2} \, $\href{http://th-dep.web.cern.ch}{CERN Theory Department},   CERN, CH-1211
  Geneva 23, Switzerland}
\end{center}

\begin{center}
{\large Abstract }
\end{center}
\begin{quote}
In this short review I  present the status of the global 21-cm signal detected by \EDGES\ in March 2018. It is organized in three parts. First, I present the \EDGES\ experiment and the fitting procedure used by the collaboration to extract the tiny 21-cm signal from large foregrounds of galactic synchrotron emission. Then, I review the physics behind the global 21-cm signature and I explain why the measured absorption feature is anomalous with respect to the predictions from standard astrophysics. I conclude with the  implications for Beyond Standard Model (BSM) physics coming from the \EDGES\ discovery. 
\end{quote}

\tableofcontents

\end{titlepage}

\section{Introduction}
In Ref.~\cite{Bowman:2018yin} the \EDGES\ collaboration has reported the discovery of a 21-cm signal in absorption between redshift 20 and 15 with an amplitude of  500 mK which is twice as large as predicted by standard cosmological computations. This signal is due to the absorption of CMB photons by the neutral atomic hydrogen and its possible detection is fundamental for astrophysics because  can open a new window onto the early phases of cosmic structure formation and give us information about the epoch of reionization soon after the formation of first stars and galaxies.  

\medskip
This short review is organized as follows. In Sec.~\ref{Sec:EDGESdetection} I will present the EDGES experiment and the fitting procedure used by the collaboration to extract the tiny  21-cm signal from large foregrounds of galactic synchrotron emission. In Sec.~\ref{Sec:21cmline} I will briefly review the physics of the 21-cm line. I will also discuss the expected properties of the InterGalactic Medium (IGM) at high-redshift from standard cosmology and I will explain why the inferred value of the spin temperature  from the \EDGES\ data is anomalous. In Sec.~\ref{Sec:BSMimplications} I will discuss the implications for Beyond Standard Model (BSM) physics coming  from this anomalous measurement and I will conclude in Sec.~\ref{Sec:Conclusions}.

\section{What did EDGES see?}\label{Sec:EDGESdetection}
The \EDGES\ experiment is a very small radio telescope, 2 meter long and 1 meter high, located in the radio quiet zone in western Australia. The equipment consists in three broad-band antennas that cover a range of frequencies from 50 to 200 MHz. The low-band antenna (operating from 50 to 100 MHz) has been designed to observe a spectral distortion in the 21-cm energy band at cosmological redshift of 20 due to the absorption of CMB photons by the IGM. However, the detection of the 21-cm signal is very challenge because of the very large foregrounds of galactic diffuse synchrotron emission. The full-sky maps of the diffuse synchrotron emission at 45 MHz and 408 MHz can be found, for example, in \cite{Guzman:2010da} and \cite{Haslam:1982zz} respectively. Before subtracting the foregrounds to the data is important to stress that: $i)$ the brightness temperature in the frequency window of \EDGES\ is always above 100 K even in region far away from the galactic center; $ii)$ the galactic synchrotron emission is spectrally smooth above 50 MHz but might need several terms to model it in a proper way as discussed in details in \cite{Bernardi:2014caa}; $iii)$ the synchrotron emission features a large spatial gradient especially in the region close to the galactic center where the activity is much larger (see e.g.~\cite{Guzman:2010da, Haslam:1982zz}). 

\medskip
Fig.~1 of~\cite{Bowman:2018yin} shows the \EDGES\ detection in terms of brightness temperature as a function of the frequency obtained by looking at high galactic latitudes.  It is evident from  panel {\bf a.} that the galactic synchrotron emission dominates the observed sky noise, yielding to an almost perfect power-law profile that decreases from about 5000 K at 50 MHz to about 1000 K at 100 MHz.  Fitting and removing the galactic synchrotron emission from the spectrum with a physically motivated 5-terms polynomial the collaboration gets the residual in panel {\bf b.}. This residual is not flat and it has a root-mean-square of 87 mK. Repeating the same exercise by adding to the 5-terms polynomial a template of the signal like the one in panel {\bf d.}, the collaboration gets the residuals in panel {\bf c.}. This new residual is now flat and the fit substantially ameliorates with a root-mean-square of only 25 mK. Adding the template to the residual the 21-cm signal is finally reported in panel {\bf d.}. Fig.~2 of Ref.~\cite{Bowman:2018yin} summarizes the detected signal obtained by using different experimental configurations. As one can see this is a signal in absorption because the brightness temperature is negative. It extends from redshift  20 to redshift 15 and it has an amplitude of $500_{-500}^{+200}$ mK at 99\% CL. The value of the plateau, centered at a frequency of 78 MHz, translating to a redshift of 17.2, is quite surprising because  is $3.8\sigma$ away from the prediction of standard cosmology. As I am going to discuss in Sec.~\ref{Sec:21cmline}, the global 21-cm signal predicted from $\Lambda$CDM can not ever be below $-230$ mK. If the measured amplitude is correct, BSM physics is required. 

\section{What is the global 21-cm signature?}\label{Sec:21cmline}
In this section I discuss in some details the basics  of the cosmological 21-cm signature. For reviews on this topic see e.g.~\cite{Pritchard:2011xb}. The 21-cm line is the triplet-to singlet hyperfine transition of atomic hydrogen ground state due to the interaction of the magnetic moments of the proton and the electron. This splitting leads to two distinct energy levels separated by  $\Delta E=5.9 \cdot 10^{-6}$ eV which corresponds to a photon wavelength $\lambda = 2\pi/\Delta E \simeq$ 21.1 cm. Cosmological redshifting brings the signature to radio frequencies of order 100 Mhz.

\smallskip
For a system in thermal equilibrium the relative population of the two spin levels of hydrogen ground state is given by
\begin{equation}\label{eq:TSdefinition}
\frac{n_{\uparrow\uparrow}}{n_{\uparrow\downarrow}} = 3 \, e^{-\Delta E/T_S} \ ,
\end{equation} 
where $T_S$ is the spin temperature. The spin temperature can be coupled to either the gas or CMB temperatures depending  on the main process that excites the line in the early Universe. Three are the relevant processes: $i)$ excitation via the absorption of CMB photons. In this case the neutral hydrogen is fully coupled with the CMB photons and therefore $T_S$ is equal to the CMB temperature $T_{\rm CMB}(z)$; $ii)$ excitation due to  collisions of the neutral atomic hydrogen. In dense environment this process couples $T_S$ to the gas temperature $T_g$; $iii)$ excitation due to the Wouthuysen-Field effect~\cite{Field:1958, Wouthuysen:1952}. This process is illustrated for example in Fig.~2 of~\cite{Pritchard:2011xb}. Ly-$\alpha$ photons, produced for example by the first stars and quasars, excite the ground state of the atomic hydrogen to the 2P level. The 2P level  re-emits Ly-$\alpha$ photons, and may enter either of the two spin states. This process, known also as Ly-$\alpha$ pumping causes an asymmetric redistribution of the electrons between the hyperfine states, decoupling the neutral hydrogen from the CMB photons. 

\smallskip
Thermal equilibrium implies a relation between the relative occupation number of the atomic hydrogen 1S states and the main processes that excite the line in the early Universe. Using Eq.~\eqref{eq:TSdefinition}, the condition of thermal equilibrium writes
\begin{equation}\label{eq:ThermalEquilibrium}
T_S^{-1} = \frac{T_{\rm CMB}^{-1}+y_C T_{\rm g}^{-1}+y_\alpha T_\alpha^{-1}}{1+y_C+y_\alpha} \ ,
\end{equation} 
where $T_\alpha$ is the Ly-$\alpha$ temperature. The coefficients $y_C$ and $y_\alpha$ are the coupling coefficients due to atomic collisions and scattering of Ly-$\alpha$ photons. If the collisions and Ly-$\alpha$ pumping are not efficient the spin temperature is exactly the CMB one. On the other hand, when the collisions and the  Ly-$\alpha$ pumping are efficient the spin temperature decouples from the CMB and it tracks one of the other two temperatures. 

\medskip
In cosmological contexts the 21-cm line has been used to probe the gas column density of the Universe. The signal depends on the optical depth  of hydrogen clouds $\tau_\nu$ along the line of sight to some source of radio background with a temperature $T_R$. Since the optical depth of the Universe to 21-cm photons is small at all relevant redshifts $z$, the differential brightness temperature averaged over  all sky is given by
\begin{equation}\label{eq:T21definition}
\begin{split}
\delta T_{b} & = \frac{T_S - T_R}{1+z} (1-e^{-\tau_\nu}) \simeq \frac{T_S - T_R}{1+z} \, \tau_\nu \\
&\simeq 27 \mbox{ mK}\, x_{\rm HI} (1+\delta_b)\left(\frac{\Omega_b h^2}{0.023}\right)\left(\frac{0.15}{\Omega_{\rm M} h^2 }\frac{1+z}{10}\right)^{\frac12} \left(\frac{\partial_r v_r}{(1+z) H(z)}\right) \left(1- \frac{T_R}{T_S} \right) \ ,
\end{split}
\end{equation} 
where $H(z)$ is the Hubble's constant, $\partial_r v_r$ is the velocity gradient along the line of sight, $\delta_b$ is the fractional over-density in baryons and $x_{\rm HI}$ is the fraction of neutral hydrogen. For redshifts between the CMB decoupling and the re-ionization,  $\partial_r v_r \sim (1+z) H(z)$, $\delta_b \ll 1$ and $x_{\rm HI} \sim 1$. 

It is evident from Eq.~\eqref{eq:T21definition} that the key quantity that sets the strength of the global 21-cm signal is the spin temperature. Only if this temperature is different from the background temperature a signal will be observable. On a more specific level, if $T_S$ is smaller than the background photon temperature we expect a signal in absorption that may be the one detected by \EDGES. Inserting the measured value of the 21-cm brightness temperature in Eq.~\eqref{eq:T21definition} and assuming that the dominant source of  radio background is the CMB ($T_R = T_{\rm CMB}(z)$) one gets $T_S \simeq 3.4$ K at $z \simeq 17.2$. 

\medskip
The most important phases of the global 21-cm signature are driven by the evolution of the spin temperature. Within standard cosmological framework ($\Lambda$CDM Cosmology) these phases are summarized as follows
\begin{itemize}
\item   Recombination ($z \simeq 1060$): the gas kinetically decouples from the CMB.  The Universe promptly becomes  neutral leaving a  population of free electrons and protons of the order of $10^{-4}$ charges per neutral atomic hydrogen. 

\item   $130 \lesssim z \lesssim 1060$: since the number density of CMB photons is much larger than the baryonic one, the compton scattering of CMB photons with the residual population of free electrons is efficient in keeping the gas in thermal equilibrium with the CMB. During this phase $T_g = T_{\rm CMB}(z) \propto (1+z)$. 

The high density of the gas leads to strong collisional coupling ($y_C \gg 1$) so that also $T_S = T_g$. Since all the temperatures are the same, the fluctuations of $T_b$ in Eq.~\eqref{eq:T21definition} are negligible and therefore we do not expect any 21-cm signal. 

\item   $40 \lesssim z \lesssim 130$: the gas thermally decouples from the CMB at $z\simeq 130$. In this phase the gas cools adiabatically so that $T_g \propto (1+z)^2$.  

The density of the gas is high enough to still make collisional coupling efficient ($y_C > 1$) so that $T_S = T_g$. Since the gas is now colder than the CMB  an early  21-cm signal in absorption is foreseen. However this is not the one detected by \EDGES\ because it takes place at higher redshift. 

\item   $z_\star \lesssim z \lesssim 40$: the density of the gas decreases and the collisional coupling is no longer efficient in keeping $T_S = T_g$ ($y_C < 1$). During this phase $T_S \simeq T_{\rm CMB}(z)$ and therefore a second period  without a 21-cm signal is expected. 

\item   $z_\alpha \lesssim z \lesssim z_\star$: light turn on. The first stars and quasars emit both Ly-$\alpha$ photons and X-rays. In general the emissivity required for the Ly-$\alpha$ coupling $y_\alpha$ is less than that for heating the gas above $T_R$. 

During this period the gas is still cooling  adiabatically and the Ly-$\alpha$ pumping couples $T_S$ to  $T_\alpha$ ($y_\alpha > 1$). Note that $T_\alpha \simeq T_g$ due to the very efficient recoil of the Ly-$\alpha$ photons in the Wouthuysen-Field effect. We therefore expect a regime where the spin temperature is coupled to cold gas so that $T_S \simeq T_g < T_{\rm CMB}(z)$. Hence, a second period with a 21-cm signal in absorption is foreseen that could be the one detected by \EDGES.  

\item   $z_h \lesssim z \lesssim z_\alpha$: by this point, heating becomes significant. The gas temperature overtakes the CMB one and therefore is expected a 21-cm signal in emission. The signal will die after the full re-ionization of the Universe because $x_{\rm HI} \equiv 0$ in Eq.~\eqref{eq:T21definition}.
 
\end{itemize}

The important point to stress is that the phases with $z>z_\star$ can be determined quite precisely. On the other hand, the epochs with $z<z_\star$ are not sharply defined because the determination of them strongly relies on the star formation history in the early Universe which is unknown. A plot of the different phases performed by assuming a benchmark fiducial history of the star formation can be found for example in Fig.~9 of \cite{Taylor:2012fx}. On the top panel is shown the evolution of the main temperatures while on the bottom the evolution of the 21-cm signal. As it is apparent, the signal has all the features briefly discussed above. Two period in absorption (one at high redshift and another one at lower redshift), one period in emission until the signal dies after the full re-ionization of the Universe.   

\medskip
The spin temperature inferred from the \EDGES\ measurement is inconsistent with any standard cosmological computations even in the  conservative case in which the gas just cools adiabatically and the Ly-$\alpha$ pumping is maximally efficient in coupling the spin temperature to the gas one. Indeed the gas, composed mainly of neutral hydrogen, thermally decouples from the CMB at $z\simeq 130$. By the redshift of the \EDGES\ measurement  the adiabatic value of the gas temperature is $T_g(z=17.2) \equiv T_{\rm CMB}(z=130) \left[(1+17.2)/(1+130)\right]^2\simeq 7$ K. Assuming that the gas is not heated up and is strongly coupled to $T_S$ via the Ly-$\alpha$ pumping, a very conservative lower value of the 21-cm signal is obtained by plugging the adiabatic value of $T_g = T_S$ in Eq.~\eqref{eq:T21definition}. At redshift 17.2 we get $\delta T_b \simeq -230$ mK more than a factor 2 away from the measured absorption feature ($-500_{-500}^{+200}$ mK at 99\% CL). 

\section{Implications for BSM}\label{Sec:BSMimplications}
 In this section I  briefly discuss the implications of the  \EDGES\ detection for BSM physics. In Sec.~\ref{subsec:explain} I  present two main approaches that have been proposed to explain the anomalous amplitude of the measured 21-cm signal. In Sec.~\ref{subsec:bounds} I review the limits on the properties of any extra source of heating in the early Universe, with a special focus on the bounds from Dark Matter (DM) annihilations.

\subsection{Explain the anomaly}\label{subsec:explain}
In order to obtained a more pronounced signal in absorption than the one predicted from standard cosmology one has to find a way to increase the ratio $T_R/T_S$ in  Eq.~\eqref{eq:T21definition}  of roughly a factor 2. Two are the main ways: $i)$ keeping $T_R = T_{\rm CMB}$, the gas is cooled even more than adiabatically so that $T_S<T_{\rm CMB}(z=130) \left[(1+17.2)/(1+130)\right]^2\simeq 7$ K; $ii)$ keeping $T_S = T_g$, a non-thermal population of photons to the Rayleigh Jeans (RJ) tail of the CMB is added so that $T_R > T_{\rm CMB}$. 

Could DM do that? In principle yes, but it cannot be a normal WIMP or axion with interactions that are too weak with the baryons. It must be something peculiar as I am going to discuss below. 

\subsubsection{Cool the gas more than adiabatically}
To get an unexpected cooling of the hydrogen gas during or prior to the so called Cosmic Dawn era one basically needs two important ingredients. The first is an abundant sector of particles in the early Universe to transfer the entropy. This is quite easy because we have the DM sector which is substantially colder than the baryonic one. Then one needs large interactions between the two sectors to exchange energy in an efficient way. This is not easy because the DM-baryons interactions have to compete with  the residual compton scattering between free electrons and photons that tends to keep the gas in thermal equilibrium with the CMB. In order to beat the compton cross section the only possibility  is to assume velocity-dependent, Rutherford-like interactions, that are strongly enhanced (${\rm d}\sigma /{\rm d} \Omega  \propto v_{\rm rel}^{-4}$). At the Cosmic Dawn $v_{\rm rel} \sim 10^{-6}$ and therefore we expect an enhancement of the cross section of 24 order of magnitude. The possibility that DM cools the gas through velocity-dependent, Rutherford-like interactions has been discussed in several works (see e.g.~\cite{Barkana:2018lgd, Munoz:2018pzp, Barkana:2018cct}).  Such interactions can only arise if the mediator mass is smaller than the typical momentum $q$ transferred  in the collisions. For DM lighter than the hydrogen $|q| \simeq 2 m_{\rm DM} v_{\rm rel} \lesssim 1$ keV. As a consequence the mediator must be very light. Two possibilities exist: $i)$ either the baryons (hydrogen and helium) are charged under the new interaction. In this case the electrons and nucleons do not screen the long-range force; $ii)$ or the baryons are neutral.  

\smallskip
The former case is ruled out. Indeed as one can see from Fig.~2 of~\cite{Barkana:2018cct}, the minimal coupling needed to fit the \EDGES\ signal is excluded by many order of magnitude by 5th force experiments and stellar cooling. 

The case in which the baryons are neutral under the new force can instead arise from either DM models  with a hidden photon that mixes with the SM photon and mediate the interactions or models under which DM is millicharged and the mediator is the visible photon itself. Ref.~\cite{Barkana:2018cct} provides the most complete analysis of these two scenarios. Assuming that the 100\% of  the DM has Rutherford-like interactions with the hydrogen, both of these possibilities are ruled out due to limits on self-interacting and millicharge DM respectively (see Fig.~3 of~\cite{Barkana:2018cct} for the hidden photon case). As a consequence, the dominant DM component cannot cool the hydrogen enough to explain the observed signal. In the case of a millicharged particles, a subcomponent of the DM around 1\% with  mass of the order of 50 MeV and  $Q\approx 10^{-4}$, may explain the \EDGES\ signal while marginally evading the limits (see Fig.~4 of~\cite{Barkana:2018cct}).

\subsubsection{New Physics in the RJ tail of the CMB}
Instead of cooling the gas through velocity-dependent, Rutherford-like DM-baryons interactions, one can also increase the population of photons around the RJ tail of the CMB. Refs.~\cite{Fraser:2018acy, Pospelov:2018kdh} introduce a class of DM models that could do that. These models are peculiar and have the following properties. First, a particle that decays into very light dark photons $A'$ is needed. If the decaying particle has a mass in the milli-eV range and substantially contributes  to the total DM energy density, the expected multiplicity of dark photons is larger than the number density of photons in the RJ tail of the CMB. Then, if the oscillation $A' \leftrightarrow A$ is permitted (for example via kinetic mixing) between the redshift of recombination and the redshift of the \EDGES\ signal a resonant conversion of $A'$ into $A$ is expected. This takes place when the mass of the dark photons $m_{A'}$ is close to the plasma mass of photons given by $m_A(z)\simeq 1.7 \cdot 10^{-14} \mbox{ eV }(1+z)^{3/2} x_e^{1/2}(z)$, where $x_e(z)$ is the fraction of free electrons after CMB decoupling. If all these requirements  are satisfied, an enhanced number density of RJ quanta is foresen making a deeper than expected 21-cm signal.

\smallskip
Fig.~4 of~\cite{Pospelov:2018kdh} summarizes  the  results considering a milli-eV axion-like particle decaying into very light soft dark radiations. On the left is shown the number ratio of converted dark photons as a function of the life-time of the decaying particle, while on the right  the parameter space in the kinetic mixing as a function of the dark photon mass plane choosing the case in which the multiplicity of converted dark photons is identical to the number density of  photons in the RJ tail of the CMB.  As one can see, there is room to modify the RJ tail of the CMB spectrum without contradicting other astrophysical or cosmological constraints. One can get a deeper  than expected 21-cm signal and explain the \EDGES\ detection for $m_{A'}$ in the range $(10^{-14} - 10^{-9})$ eV with non-vanishing kinetic mixing $\epsilon \sim 10^{-7}$ with the visible photon.  The life-time of the milli-eV axion-like particle is in the range $(10-10^5)\, \tau_U$. This significant parameter space can by mostly probed by the next generation of CMB experiments (e.g.~PIXIE/PRISM).

\subsection{Bounds on new physics}\label{subsec:bounds}
The unexpected low value of the spin temperature inferred by \EDGES\ gives us the great opportunity to put  severe and stringent limits on the properties of any extra source of heating. In particular, new constraints on DM annihilation and decay can be set by using the information on the thermal history provided by the 21-cm measurement (see e.g.~\cite{DAmico:2018sxd, Liu:2018uzy, Mitridate:2018iag}). 

\subsubsection{Bounds on DM annihilations}
\begin{figure}[t!]
    \centering
    \includegraphics[width=0.495\textwidth]{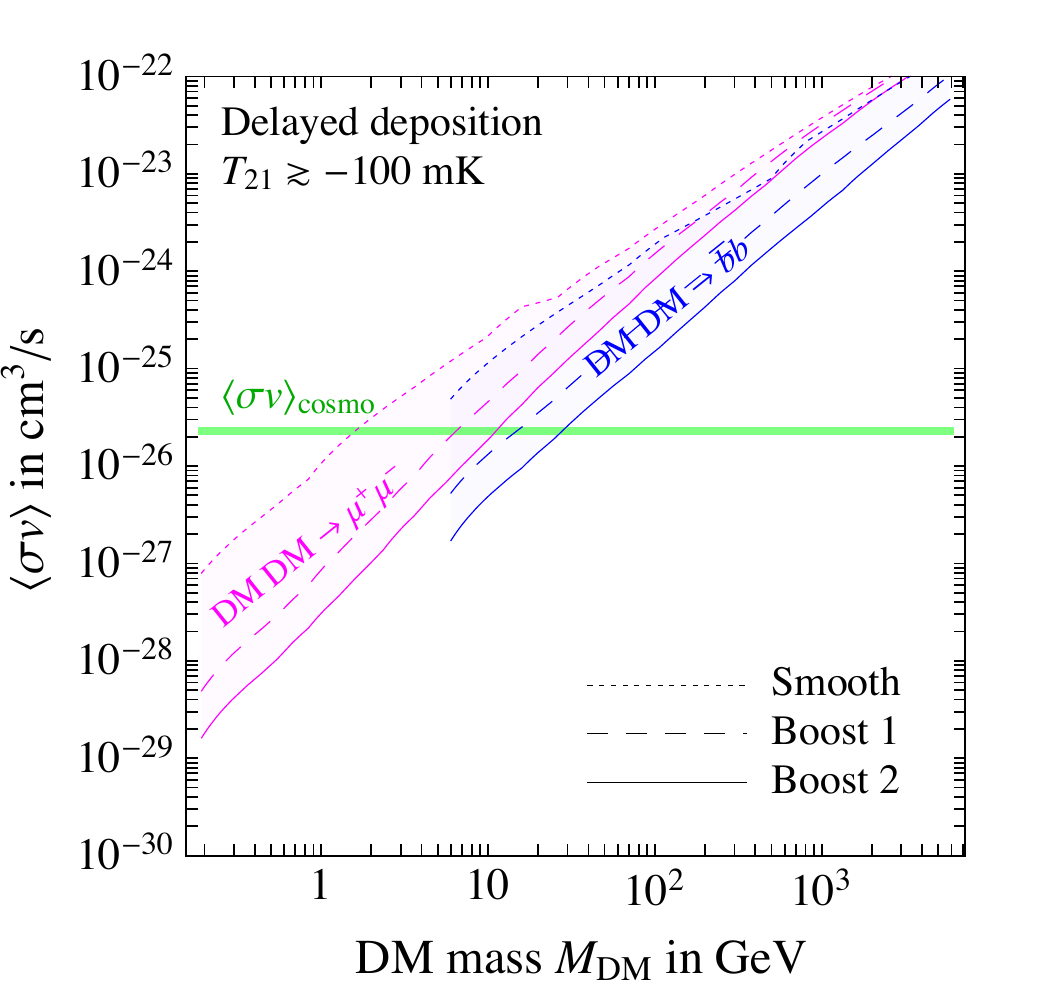}\
    \includegraphics[width=0.495\textwidth]{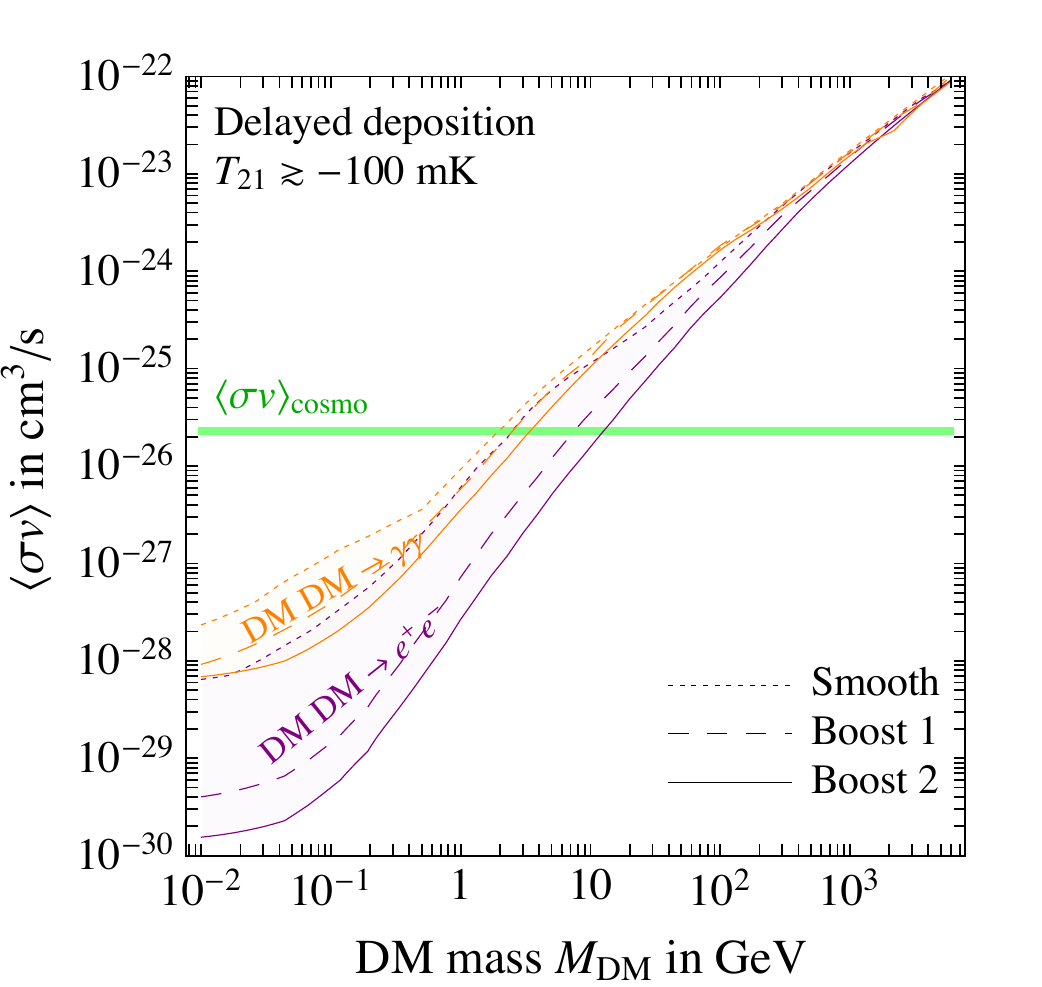}\
    \caption{\em \label{fig:fsigmav} Bounds on DM annihilation cross sections $\langle \sigma v \rangle$ as a function of the DM mass $M_{\rm DM}$ from Ref.~\cite{DAmico:2018sxd}. We demand that  the brightness temperature fluctuations $\delta T_b \equiv T_{21} \gtrsim 100 \, \mathrm{mK}$. We report the results for two different cosmological boost factors, as well as ignoring DM clustering. Left (right) panel  shows DM annihilating into bottom quarks or muons (photons or electrons).}
\end{figure}
DM annihilation products can considerably heat the gas, suppressing the observed absorption feature, even erasing it if DM heating is too large. Therefore, the observation of an absorption feature in the 21-cm spectrum implies bounds on DM annihilations. Annihilating DM can heat the IGM in two main ways (see e.g.~\cite{Cirelli:2009bb}). First, the annihilations increase the number of free electrons and positrons after the CMB kinetic decoupling. The extra population of $e^\pm$ makes compton scattering more efficient in keeping the gas in thermal equilibrium with the CMB.  A delay in the hydrogen/CMB thermal decoupling implies an higher $T_g$ at lower redshift since the gas had less time to cool adiabatically. The second more efficient way to heat the gas is via  direct injection of the energy into the plasma.  DM annihilations produce indeed stable SM products like electrons, positrons and photons that initiate an electromagnetic shower in the early Universe that directly heat the gas. A higher $T_g$ will result in a modification of $\delta T_b$ as discussed, for example, in~\cite{Valdes:2012zv, Lopez-Honorez:2016sur}. 

\smallskip
In Ref.~\cite{DAmico:2018sxd} we have derived for the first time limits on DM annihilations from the new \EDGES\ measurement. Without entering in the details of the paper, I summarize here the strategy we choose to set the limits. We do not try to explain the signal which is lower than expected. DM annihilations, as I briefly discussed above, increase  $T_g$ and in turn tend to erase the 21-cm signal in absorption. As a consequence, we can put bounds, but we cannot use the measured value of the spin temperature by \EDGES\ because is already $3.8\sigma$ away from the predictions from standard cosmology.  Our strategy is to assume standard evolution in the very conservative scenario in which the gas just cool adiabatically and the Ly$\alpha$-pumping is so efficient that recouple the spin temperature to the gas one around the redshift range of \EDGES. This gives the strongest absorption of the order of $-230$ mK as discussed in Sec.~\ref{Sec:21cmline}. Then, by adding DM annihilations, we require that it does not erase the 21-cm above $-100$ mK. 

Fig.~\ref{fig:fsigmav} shows the limits on DM annihilation cross sections $\langle \sigma v \rangle$ as a function of the DM mass $M_{\rm DM}$ for some representative primary channels. We consider two different cosmological boost factors (Boost 1 (dashed line) and 2 (solid line)), as well as the weaker bound obtained by ignoring DM clustering (dotted line). We compute the energy injection and deposition into the IGM by convolving the primary spectra provided in~\cite{Cirelli:2010xx} with the  delayed transfer functions of~\cite{Slatyer:2015jla}. The effects due to the boost factor vary with DM mass and primary annihilation channel. This can be understood because energetic electrons and photons  from DM annihilations deposit in the IGM a relevant amount of their energy only after some time. The efficiency of the energy deposition depends a lot on the primary spectra.  For hadronic channels (such as quarks, $\tau^+\tau^-, W^+ W^-, ZZ$ and $hh$) energy deposition is well approximated as instantaneous. For pure electromagnetic  channels (such as $\gamma\gamma$ and $e^+ e^-$) the deposition is delayed. This explain why the limits on the right panel (pure electromagnetic channels) become almost independent on the cosmological boost factor for $M_{\rm DM}> 1$ GeV. The 21-cm limits  are comparable to those from the CMB~\cite{Ade:2015xua}, and to limits from indirect DM searches~\cite{Ackermann:2015zua}. 

\section{Conclusions}\label{Sec:Conclusions}
Thanks to the new \EDGES\ discovery we have started  to probe the Universe in the 21-cm wavelength. The \EDGES\ detection, if confirmed, is fundamental for astrophysics and perhaps for particle physics because   can open a new window onto the early phases of cosmic structure formation and give us information about the epoch of reionization soon after the formation of first stars and galaxies. The measured  21-cm signal is in absorption.  It extends from redshift  20 to redshift 15 and it has an amplitude of $500_{-500}^{+200}$ mK at 99\% CL. The value of the plateau, centered at a frequency of 78 MHz, translating to a redshift of 17.2, is quite surprising because  is $3.8\sigma$ away from the prediction of standard cosmology around $-230$ mK. 

\smallskip
In this short review I have presented two possible approaches that have been proposed to explain the anomalous value of $\delta T_b$ that only select few peculiar BSM models. Furthermore, since the value of the spin temperature inferred by \EDGES\ is tiny, one has the great opportunity to put severe and stringent limits on the properties of any extra source of IGM heating. In Ref.~\cite{DAmico:2018sxd} we have derived for the first time limits on DM annihilations. These limits are completive and in some cases the most stringent bounds in the literature. This is just the beginning, stay tuned for further developments in the field of cosmological 21-cm.


\end{document}